\documentstyle[12pt]{article}

\begin{document}

\begin{center}
{\LARGE {\bf Consistent analytical approach for the quasiclassical radial
dipole matrix elements}}$\footnote{
Published in {\it J. Phys. B: At. Mol. Opt. Phys.} {\bf 28 }(1995) 4963-4971.
}${\LARGE {\bf \ }}
\end{center}

\thispagestyle{empty}

\vspace{1cm}

{\qquad \quad B. Kaulakys\footnote{
E-mail: kaulakys@itpa.lt}} \vspace{0.3cm}

{\hfill
\parbox[t]{12cm}{
{\small Institute of Theoretical Physics and Astronomy, A. Go\v stauto 12,
2600 Vilnius, Lithuania} \vspace{1cm}

Received

\vspace{1cm}
Short title: {\bf Quasiclassical dipole matrix elements}
\vspace{1cm}

{\bf Abstract.} A consistent analytical approach for calculation of the 
quasiclassical radial dipole matrix elements in the momentum and coordinate 
representations is presented. Very simple but relatively precise expressions 
for the matrix elements are derived in both representations. All analytical 
expressions contain only one special function -- the Anger function and its 
derivative. They generalise and increase the accuracy of some known 
quasiclassical expressions. The small difference between the two forms of 
the expressions for the dipole matrix elements indicates to
the applicability of the simple expressions given by the consistent
quasiclassical approach even for low atomic states.}} \vspace{1cm}

 \vspace{1cm}

\newpage
\setcounter{page}{1}

{\bf 1. Introduction }\vspace{0.5cm} \\ Calculation of transition
probabilities, oscillator strengths and dipole matrix elements for the
atomic transitions is stimulated of investigations in spectroscopy, plasma
physics, chaotic dynamics of the nonlinear systems and other fundamental and
applied fields. The accurate calculations of these atomic characteristics
are very time consuming and in some cases involve certain difficulties,
especially for transitions between high states. Therefore, recently new
attention has been paid to quasiclassical calculations of the dipole matrix
elements and oscillator strengths for atomic $nl\rightarrow n^{\prime
}l^{\prime }$ transitions (Heim {\it et al} 1989, Delone {\it et al} 1989,
1994, Kaulakys 1991, D'yachkov and Pankratov 1991, 1994, Pankratov and
Meyer-ter Vehn 1992 and Nana {\it et al }1995). In the review paper by
Delone {\it et al} (1994) the set of analytical formulae for the dipole
matrix elements between quasiclassical states is presented. However, this
review paper does not reflect the results of the last five-year period by
Heim {\it et al} (1989), Kaulakys (1991), D'yachkov and Pankratov (1991,
1994) Pankratov and Meyer-ter Vehn (1992) and others and contains some
inconsistencies. As a result the main formulae of the paper contain terms
with erroneous sign and do not agree with the exact results, e.g. when $
s=\nu ^{\prime }-\nu $ is an integer number and when $s\rightarrow 0$ 
with $ \nu $ and $\nu ^{\prime }$ being the effective principal quantum
number of the initial and final state, respectively.

Here we present a consistent analytical approach for the quasiclassical
dipole coupling of the electron with the electromagnetic field, taking into
account peculiarity of the radial quasiclassical matrix elements: radial
wave functions of the initial and final states for transition with $\Delta
l\neq 0$ correspond to the different effective Hamiltonians and, therefore,
we can not use the usual correspondence between the quasiclassical matrix
elements and the Fourier components of the appropriate classical variable
(see e.g. Landau and Lifshitz 1985). So we should start from the definition
of the quantum matrix elements using the quasiclassical radial wave
functions.

The direct way of coupling a radiation field to the electron Hamiltonian is
through the ${\bf A}\cdot {\bf p}$ interaction (see, e.g. Bethe and Salpeter
1957 and Landau and Lifshitz 1985) where ${\bf A}$ is the vector potential
of the electromagnetic field and ${\bf p=-}i\hbar {\bf \nabla }$ is the
momentum operator. When the radiation wavelength is long compared with
atomic dimension, as is the case for optical or microwave transitions, the
variation of the vector potential within the atom can be neglected. In this
case the electric field is also uniform over the extent of the atom. So that
the electric dipole moment is the only atomic multipole coupled to the field
in the multipolar approximation to the interaction operator and the
interaction of the electron with the field may also be expressed through the
${\bf E\cdot r}$ term, where ${\bf E}$ is the electric field strength and $
{\bf r}$ is the electron's coordinate. As a matter of fact, the two forms of
the interaction Hamiltonian in the dipole approximation are equivalent due
to the gauge invariance of the field.

Usually one calculates the radial dipole matrix elements in the coordinate
form (see Heim {\it et al} 1989, Delone {\it et al} 1989, 1994, Kaulakys
1991, D'yachkov and Pankratov 1991, 1994, Pankratov and Meyer-ter Vehn 1992,
Nana {\it et al }1995 and references therein). For the precise wave
functions of the $\mid a\rangle $ and $\mid b\rangle $ states the relation
between the matrix elements of ${\bf p}$ and ${\bf r}$ operators
$$
{\bf p}_{a,b}=-im\omega _{b,a}{\bf r}_{a,b}\eqno(1)
$$
holds. Here $m$ and $\omega _{b,a}=(E_b-E_a)/\hbar $ are the electron mass
and the angular frequency of the corresponding transition, respectively.
However, when one uses approximate wave functions in the evaluation of the
matrix elements, the length and velocity forms in general yield different
results and the relationship (1) holds only roughly or even fails (see, e.g.
Crossley 1969). If asymptotically correct wave functions are used, the 
{\it r}-form of the dipole matrix elements is preferable as it stresses the
contributions to the integral from the large $r$ region. On the other hand,
for the wave functions correct at small and medium $r$ the velocity form
should be used as it puts more weight on the integral at small and medium 
$r$. The quasiclassical wave functions in the classically allowed region of
motion are not correct asymptotically as well as for very small $r$ but are
relatively accurate for the medium $r$ between the two turning points of the
classical orbit. Therefore, it is likely that the velocity form of the
quasiclassical matrix elements is as much (or, maybe, even more) accurate as
the length form. \vspace{1cm} \\ {\bf 2. Radial dipole matrix elements for
any spherical potential }\vspace{0.5cm} \\Calculation of the angular part of
the matrix elements is a simple problem and, therefore, we restrict
ourselves to the radial part. The radial dipole matrix element in the
momentum representation is given by \footnote{
Further in the paper we will use the atomic units, $\hbar =m=e=1$.}
$$
D_{nl}^{n^{\prime }l\pm 1}=\frac 1\omega \int P_{nl}\left( r\right) \left[
\frac{dP_{n^{\prime }l\pm 1}\left( r\right) }{dr}\pm l_{\max }\frac{
P_{n^{\prime }l\pm 1}\left( r\right) }r\right] dr\eqno(2)
$$
were $P_{nl}(r)=rR_{nl}(r)$ is the solution of the radial Schr\"odinger
equation, $\omega =E_{n^{\prime }l^{\prime }}-E_{nl}$ is the transition
frequency and $l_{\max }=\max (l,l\pm 1)$. The quasiclassical radial wave
function $P_{nl}$ can be expressed as
$$
P_{nl}=\frac 2{\sqrt{Tv_r(r)}}\cos \Phi _{nl}(r)\eqno(3)
$$
in the classically allowed region of motion and some exponentially
decreasing function outside this region. Here $T$ is the period of classical
rotation, the radial velocity of the electron $v_r$ is given by
$$
v_r(r)=\left[ 2E_{nl}-2U(r)-\frac{(l+\frac 12)^2}{r^2}\right] ^{1/2}\eqno(4)
$$
and the phase $\Phi _{nl}$ is defined as
$$
\Phi _{nl}(r)=\int_{r_1}^rv_r(r)dr-\frac 14\pi \eqno(5)
$$
with $U(r)$ being the effective potential which defines the motion of the
Rydberg electron and roots of the radial velocity $v_r(r),$ $r_1$ and $r_2$,
are the two classical turning points given by $v_r(r_1)=v(r_2)=0.$

Further calculations of the matrix elements (2) are similar to those by Heim
{\it et al} (1989) and Kaulakys (1991). Substituting equation (3) into
equation (2) and neglecting the rapidly oscillating sine and cosine of the
sum of the phases $\Phi _{nl}$ and $\Phi _{n^{\prime }l^{\prime }}$ one
finally finds that
$$
D_{nl}^{n^{\prime }l\pm 1}=\frac 2{\omega \sqrt{TT^{\prime }}
}\int\limits_0^{T_c/2}\left[ -\dot r\sin \Delta \Phi (t)\pm r\dot \varphi
\cos \Delta \Phi (t)\right] dt
$$
$$
=\frac 2{\omega \sqrt{TT^{\prime }}}\int\limits_0^{T_c/2}[-\dot x\sin \omega
t\pm \dot y\cos \omega t]dt\eqno(6)
$$
where $\varphi $ is the polar angle while $x$ and $y$ are Cartesian
coordinates of the electron, points denote the derivatives with respect to
time and $T_c$ is some mean period to be defined later. In the derivation of
equation (6) we have used the fact that $r^{-1}l_{\max }=r^{-1}\overline{
(l+\frac 12)}=r^{-1}(l+l^{\prime }+1)/2=r^{-1}(r^2\dot \varphi )=r\dot
\varphi $ and that, according to equations (4) and (5), linear expansion of
the difference of the phases in powers of $\omega $ and $\Delta l=l^{\prime
}-l$ is
$$
\Delta \Phi \equiv \Phi _{n^{\prime }l^{\prime }}-\Phi _{nl}\simeq \omega
t-\Delta l\varphi +....\eqno(7)
$$

Integration of equation (6) by parts yields to the $r$-form of the radial
dipole matrix element
$$
R_{nl}^{n^{\prime }l\pm 1}\equiv \langle nl\mid r\mid n^{\prime }l\pm
1\rangle =\frac 2{\sqrt{TT^{\prime }}}\int\limits_0^{T_c/2}r(t)\cos \Delta
\Phi (t)]dt
$$
$$
=\frac 2{\sqrt{TT^{\prime }}}\int\limits_0^{T_c/2}[x(t)\cos \omega t\pm
y(t)\sin \omega t]dt.\eqno(8)
$$
Here we have used the quasiclassical quantisation conditions (see Kaulakys
1991 for details) and facts, that
$$
\Delta \Phi (0)\simeq 0,\quad\Delta \Phi (T_c/2)=\Delta n_r\pi ,\quad
\omega T_c/2=\Delta n\pi ,
$$
$$
r(0)=x(0)=r_1,\quad y(0)=y(T_c/2)=0,\quad\varphi (0)=0,\quad\varphi
(T_c/2)=\pi \eqno(9)
$$
where $n_r=n-l-1$ is the radial quantum number and $\Delta n$ and $\Delta
n_r $ are integers. Note, that equation (8) may also be derived from the
definition of the radial dipole matrix element in the coordinate
representation and using expression (3) for the quasiclassical radial wave
function (see Kaulakys 1991).

It follows from equations (6)--(8) that only for $\Delta l=0$ the
quasiclassical matrix elements of the module of the radius vector $r$
coincide with the Fourier components of the classical variable $r(t)$ and
may be expressed through the Fourier components of the classical variable $
\dot r(t)$, i.e.
$$
\langle nl\mid r\mid n^{\prime }l\rangle =\frac 2{\sqrt{TT^{\prime }}
}\int\limits_0^{T_c/2}r(t)\cos \omega tdt=\frac 1{\sqrt{TT^{\prime }}}\oint
r(t)e^{-i\omega t}dt
$$
$$
=-\frac 2{\omega \sqrt{TT^{\prime }}}\int\limits_0^{T_c/2}\dot r(t)\sin
\omega tdt=\frac{-i}{\omega \sqrt{TT^{\prime }}}\oint \dot r(t)e^{-i\omega
t}dt.\eqno(10)
$$

For the dipole transitions with $\Delta l=\pm 1$ radial wave functions $
P_{nl}$ and $P_{n^{\prime }l^{\prime }}$ of the initial and final states are
solutions of the radial Schr\"odinger equation with different effective
potentials $V_{eff}(r)=U(r)+(l+\frac 12)^2/2r^2$ and $V_{eff}^{\prime
}(r)=U(r)+(l^{\prime }+\frac 12)^2/2r^2$, respectively. This results to the
additional term in the phase difference (7) and second term in equations (6)
and (8).

Note, that the radial dipole matrix elements $D_{nl}^{n^{\prime }l\pm 1}$
and $R_{nl}^{n^{\prime }l\pm 1}$ according to equations (6) and (8) may also
be expressed as
$$
D_{nl}^{n^{\prime }l\pm 1}=\frac{-i}{\omega \sqrt{TT^{\prime }}}\oint \left[
\dot x(t)\pm \dot y(t)\right] e^{-i\omega t}dt\eqno(11)
$$
and
$$
R_{nl}^{n^{\prime }l\pm 1}=\frac 1{\sqrt{TT^{\prime }}}\oint \left[ x(t)\pm
y(t)\right] e^{-i\omega t}dt.\eqno(12)
$$

It is of interest to note the connection between the expressions for the
dipole matrix elements (6) -- (10) and the energy change of the classical
atom in a monochromatic field (see Gontis and Kaulakys 1987 and Kaulakys
1991). The mapping equations of motion for the classical hydrogen atom in an
oscillating electric field derived on the bases of the velocity form of
interaction are suitable for investigation of transition to chaotic
behaviour and ionisation of Rydberg atoms even in the low frequency field
when the strength of the external field is comparable with the Coulomb field
of the atom (Kaulakys and Vilutis 1995).

Until now we did not use the explicit form of the potential $U(r).$
Therefore, equations (1)--(12) are valid for any spherical potential.
Further we will apply this theory for the non-hydrogenic atoms. \vspace{1cm}
\\ {\bf 3. The non-hydrogenic atom }\vspace{0.5cm} \\ The potential $U(r)$
which defines the motion of the Rydberg electron of the non-hydrogenic atom
or ion may be expressed as a sum of two terms ---the Coulomb potential $-Z/r$
of the ion core with charge $Z=Z_{ion}+1$ and the perturbation potential $
\Delta U(r)$ due to the deviation from the Coulomb approximation, i.e.

$$
U(r)=-Z/r+\Delta U(r).\eqno(13)
$$
For the non-hydrogenic atom the perturbation potential $\Delta U(r)$ is
short-range and results to the non-Coulomb scattering phase $\delta _l$
related with the quantum defect $\mu _l$ by the relationship $\delta _l=\pi
\mu _l$.The energy $E_{nl}$ of the $\mid nl\rangle $-state is related to the
effective principal quantum number $\nu $ and the quantum defect $\mu _l$ in
the usual way
$$
E_{nl}=-Z^2/2\nu ^2,\quad\nu =n-\mu _l.\eqno(14)
$$

Significant contribution to the integrals (6) and (8) arise from the regions
with relatively large $r$ where the potential $U\left( r\right) $ is well
represented by the Coulomb potential $V\left( r\right) =-Z/r$. The
additional potential $\Delta U(r)$ results mainly to the non-Coulomb
scattering phaseshift $\delta _l$ (see Kaulakys 1991). Thus, the phase (5)
and the phase difference (7) in the region of the main contribution to the
dipole matrix element may be represented as
$$
\Phi _{nl}(r)=\int_{r_1^c}^rv_r^c(r)dr+\delta _l-\frac 14\pi \eqno(15)
$$
and
$$
\Delta \Phi \simeq \omega t-\Delta l\varphi \simeq \Delta +\omega t_c-
\Delta l\varphi _c\eqno(16)
$$
where $v_r^c$ is the radial velocity for the Coulomb potential, $\Delta
=\delta _{l^{\prime }}-\delta _l$ and it is convenient to introduce the
parametric equations of motion for the Coulomb potential
$$
r=\left( \nu _c^2/Z\right) \left( 1-e\cos \xi \right) ,\quad t_c=\left(
\nu _c^3/Z^2\right) \left( \xi -e\sin \xi \right)
$$
$$
x=\left( \nu _c^2/Z\right) \left( \cos \xi -e\right) ,\quad y=\left( \nu
_c^2/Z\right) \left( 1-e^2\right) ^{1/2}\sin \xi ,\eqno(17)
$$
$$
\varphi _c=\arccos \left( \frac{\cos \xi -e}{1-e\cos \xi }\right) ,\quad
e=\left[ 1-\left( \frac{l+l^{\prime }+1}{2\nu _c}\right) ^2\right] ^{1/2}.
$$
Here $e$ denotes the eccentricity of the classical orbit with the period 
$ T_c=2\pi \nu _c^3/Z^2$ and the turning points $r_{1,2}^c=\left( \nu
_c^2/Z\right) \left( 1\mp e\right) .$ The centered effective principal
quantum number $\nu _c$ is defined from the requirement that the phase
difference (16) at the turning point $r_2$ has to be in consistence with 
the quasiclassical quantisation conditions, i.e.
$$
\Delta \Phi \left( r_2^c\right) =\Delta +\frac 12\omega T_c-\Delta l\pi
=\Delta n_r\pi ,\eqno(18)
$$
which, together with the relationship $\Delta n_r\pi =\Delta \nu \pi +
\Delta -\Delta l\pi $, results to the expression (see also D'yachkov 
and Pankratov 1991, 1994)
$$
\nu _c^3=\frac{Z^2\Delta \nu }\omega =\frac{2\left( \nu \nu ^{\prime
}\right) ^2}{\nu +\nu ^{\prime }}.\eqno(19)
$$

More precisely, the non-Coulomb phase shifts $\delta _l$ and $\delta
_{l^{\prime }}$ and, consequently the phase difference $\Delta =\delta
_{l^{\prime }}-\delta _l$ are some functions of the $r$ and time $t$ (see
Kaulakys 1991). However, for the non-hydrogenic atoms the phase difference $
\Delta \left( r\right) $ increases in the region $r\simeq r_1\simeq r_1^c$
from $\Delta =0$ to $\Delta =\delta _{l^{\prime }}-\delta _l=const$ in the
very narrow interval of the coordinate $r$, while the main contributions to
the radial integrals (6) and (8) occur at large distances, $r\sim \nu ^2.$
Thus, in equations (6) and (8) the phase $\omega t$ may be replaced by the
expression $\Delta +\omega t_c$, while $\varphi \simeq \varphi _c.$Keeping
this in mind and substituting equations (16)--(19) into equations (6) and
(8) we have
$$
D_{nl}^{n^{\prime }l\pm 1}=\left( -1\right) ^{\Delta n}\frac{\nu _c^5}{
Z\left( \nu \nu ^{\prime }\right) ^{3/2}}D_p^{\pm }\left( e,s\right) ,
\eqno(20)
$$
$$
D_p^{\pm }=\frac 1s\left[ {\bf J}_{-s}^{\prime }\left( es\right) \pm \sqrt{
e^{-2}-1}\left( {\bf J}_{-s}\left( es\right) -\frac{\sin \pi s}{\pi s}
\right) \right] ,\eqno(21)
$$
$$
R_{nl}^{n^{\prime }l\pm 1}=\left( -1\right) ^{\Delta n}\frac{\nu _c^5}{
Z\left( \nu \nu ^{\prime }\right) ^{3/2}}D_r^{\pm }\left( e,s\right) ,
\eqno(22)
$$
$$
D_r^{\pm }=D_p^{\pm }+\frac{1-e}{\pi s}\sin \pi s,\quad s=\Delta \nu =\nu
^{\prime }-\nu .\eqno(23)
$$
Here ${\bf J}_{-s}\left( z\right) $ and ${\bf J}_{-s}^{\prime }\left(
z\right) $ are the Anger function defined as
$$
{\bf J}_{-s}\left( z\right) =\frac 1\pi \int_0^\pi \cos \left( s\xi +z\sin
\xi \right) d\xi \eqno(24)
$$
and its derivative with respect to the argument $z$, respectively. Note to
the properties of the Anger function: ${\bf J}_{-s}\left( z\right) ={\bf J}
_s\left( -z\right) $ and ${\bf J}_{-s}^{^{\prime }}\left( z\right) =-{\bf J}
_s^{^{\prime }}\left( -z\right) $ which result to the symmetry of the matrix
elements (21) and (23)

$$
D^{\pm }(e,-s)=D^{\mp }(e,s).\eqno(23')
$$

To the best of our knowledge equations (20) and (21) are derived for the
first time while equations (22) and (23) within the accuracy of factor $
\left( \nu _c/\sqrt{\nu \nu ^{\prime }}\right) ^5$ coincide with the
equation (16) in the paper by Kaulakys (1991) and are close to the
corresponding expressions given by D'yachkov and Pankratov (1991) and by
Pankratov and Meyer-ter-Vehn (1992) (see also Nana {\it et al} 1995). Note,
that for the first time the dipole matrix elements in the coordinate
representation have been expressed{\it \ through the Anger function and its
derivative} (however, with some erroneous signs) by Davydkin and Zon (1981).

In the derivation of equations (22) and (23) one integrates equation (8) by
parts using the approximate expression $\omega t\simeq \Delta +\omega t_c$.
This yields to equation (6) with the additional term
$$
R_{nl}^{n^{\prime }l\pm 1}-D_{nl}^{n^{\prime }l\pm 1}=-\frac{2r_1^c}{\omega
\sqrt{TT^{\prime }}}\sin \Delta =\left( -1\right) ^{\Delta n}\frac{\nu
_c^5\left( 1-e\right) \sin \pi s}{Z\left( \nu \nu ^{\prime }\right)
^{3/2}\pi s}.\eqno(25)
$$
Thus, the difference between the quasiclassical dipole matrix elements in
the {\it r}-and {\it p}-forms, the second term in equation (23), results
from the replacement of the electron's motion in the effective potential $
U(r)$ by the motion in the Coulomb potential with the additional phase $
\delta _l$. Therefore, the quasiclassical radial dipole matrix element in
the coordinate representation (22)--(23) may contain some additional
inaccuracy. On the other hand, the difference (25) between two forms of the
dipole matrix elements may be as a criterion of the exactness of the
quasiclassical approximation. As a rule, the additional term (25) is small
because of the small factor $\left( 1-e\right) $ for states with low $l$ and
of the small factor $\sin \pi s$ for states with larger $l$ but small
quantum defects and, consequently, with $s=\Delta \nu \simeq \Delta n$ close
to the integer. In expansion of the dipole matrix elements in terms of $
\alpha =\sqrt{1-e^2}=\left( l+l^{\prime }+1\right) /2\nu _c$ (see Kaulakys
1991 for analogy)
$$
D_p^{\pm }=\frac 1s\left[ {\bf J}_{-s}^{\prime }\left( s\right) \pm \alpha
\left( {\bf J}_{-s}\left( s\right) -\frac{\sin \pi s}{\pi s}\right) +\frac{
\alpha ^2}2\left( {\bf J}_{-s}^{\prime }\left( s\right) +\frac{2\sin \pi s}{
\pi s}\right) \right] \eqno(26)
$$
$$
D_r^{\pm }=D_p^{\pm }+\frac{\alpha ^2}2\frac{\sin \pi s}{\pi s}\eqno(27)
$$
this additional term makes up less than half of the third, proportional to
the $\alpha ^2$, expansion term. This indicates to the relatively high (up
to some percents) accuracy of the very simple quasiclassical approximation
(20)--(23) for the dipole matrix elements even for the low atomic states.
The extensive analysis of such approach and comparison with the numerical
Hartree-Fock calculations will be presented elsewhere. Here we will present
only the limiting forms of the dipole matrix elements. \vspace{0.5cm} \\
{\bf 4. Special cases of the parameters} \vspace{0.5cm} \\ Using the
expansions of the functions ${\bf J}_{-s}\left( es\right) $ and ${\bf J}
_{-s}^{\prime }\left( es\right) $ in powers of $s$ (Kaulakys 1991)
$$
{\bf J}_{-s}\left( es\right) \simeq 1-\left( \frac 16\pi ^2+e+\frac
14e^2\right) s^2
$$
$$
{\bf J}_{-s}^{\prime }\left( es\right) \simeq -\left( 1+\frac 12e\right) s,
\quad s\ll 1\eqno(28)
$$
we have from equations (20)--(23)
$$
D_{nl}^{n^{\prime }l\pm 1}=\left( -1\right) ^{\Delta n+1}\frac{\nu ^2}
Z\left( 1+\frac 12e\right) \eqno(29)
$$
$$
R_{nl}^{n^{\prime }l\pm 1}=\left( -1\right) ^{\Delta n+1}\frac 32\frac{\nu ^2
}Ze\quad s\ll 1.\eqno(30)
$$
For hydrogenic atom with $n^{\prime }=n$ equation (30) results to the exact
expression
$$
R_{nl}^{n^{\prime }l\pm 1}=-\frac 32\frac{n^2}Ze.\eqno(31)
$$
Substitution of the asymptotic, $s\gg 1$, forms of functions ${\bf J}
_{-s}\left( s\right) $ and ${\bf J}_{-s}^{\prime }\left( s\right) $
(Kaulakys 1991)
$$
{\bf J}_{-s}\left( s\right) =\frac{2a}{\sqrt{3}s^{1/3}}\cos \left( \pi
s-\frac 16\pi \right) \quad a\simeq 0.447
$$
$$
{\bf J}_{-s}^{\prime }\left( s\right) =\frac{2b}{\sqrt{3}s^{2/3}}\cos \left(
\pi s+\frac 16\pi \right) \quad b\simeq 0.411\eqno(32)
$$
into equations (26) and (27) yields
$$
D_{p,r}^{\pm }\simeq \frac{2b}{\sqrt{3}s^{5/3}}\cos \left( \pi s+\frac 16\pi
\right) \pm \frac{2\alpha a}{\sqrt{3}s^{4/3}}\cos \left( \pi s-\frac 16\pi
\right) .\eqno(33)
$$
From equation (33) the Bethe rule can be seen: principal and orbital quantum
numbers change prevailing in the same direction but only when $\alpha $ and $
l$ are not small and $\cos \left( \pi s+\frac 16\pi \right) $ and $\cos
\left( \pi s-\frac 16\pi \right) $ are of the same sign, e.g. when $s$ is
close to the integer. In the later case ($s=\Delta n$) we have the improved
result of Goreslavsky {\it et al} (1982) for the removed states
$$
D_{nl}^{n^{\prime }l\pm 1}\simeq R_{nl}^{n^{\prime }l\pm 1}\simeq \frac{\nu
_c^5}{Z\left( \nu \nu ^{\prime }\right) ^{3/2}}\left( \frac b{\left( \Delta
n\right) ^{5/3}}\pm \frac{\alpha a}{\left( \Delta n\right) ^{4/3}}\right)
$$
$$
=Z^{7/3}\left[ b\pm \alpha a\left( \Delta n\right) ^{1/3}\right] /\omega
^{5/3}\left( \nu \nu ^{\prime }\right) ^{3/2}\quad \Delta n\gg 1.\eqno(34)
$$
On the other hand, for large $s=\Delta n$ the Anger function and its
derivative may be expressed through the Airy function and its derivative or
through the McDonald functions. As a result we have from equations
(20)--(23)
$$
D_{p,r}^{\pm }=\left( -1\right) ^{\Delta n}\frac{\sqrt{2}\zeta ^{3/4}\left(
1-e^2\right) ^{1/4}}{\pi \sqrt{3}e\Delta n}\left[ K_{2/3}\left( \frac
23s\zeta ^{3/2}\right) \pm K_{1/3}\left( \frac 23s\zeta ^{3/2}\right)
\right] \eqno(35)
$$
where
$$
\frac 23\zeta ^{3/2}=\ln \frac{1+\sqrt{1-e^2}}e-\sqrt{1-e^2}\eqno(36)
$$
and $K_\nu \left( z\right) $ are the McDonald functions.

For $1-e^2=\alpha ^2\ll 1$ it yields from equations (35) and (36)
$$
D_{p,r}^{\pm }=\left( -1\right) ^{\Delta n}\frac{\alpha ^2}{\pi \sqrt{3}
e\Delta n}\left[ K_{2/3}\left( \frac 13\alpha ^3\Delta n\right) \pm
K_{1/3}\left( \frac 13\alpha ^3\Delta n\right) \right] .\eqno(37)
$$

In the limit $\frac 13\alpha ^3\Delta n\ll 1$ equation (37) results to the
expression (34).

The dipole matrix elements for transitions between states with the large, $
l\sim n$, orbital quantum numbers, as follows from equations (20)--(23) or
(35) and (36) when $e\rightarrow 0$, are exponentially small. Moreover, the
Bethe rule in this case is enhanced: the transitions with the change of
principal and orbital quantum numbers in the opposite directions are
strongly suppressed in comparison with transitions, when $n$ and $l$ change
in the same direction.

Thus, the very simple expressions (20)--(23) cover all known quasiclassical
non-relativistic results for the dipole matrix elements. They generalize and
increase the accuracy of some earlier derived expressions. Extension of the
present approach to the continuum states is rather straightforward (see
Kaulakys 1991 for analysis in the {\it r}-representation). \vspace{1cm} \\
{\bf 5. Conclusions} \vspace{0.5cm} \\ Consistent analytical approach for
calculation of the quasiclassical radial dipole matrix elements in the
momentum and coordinate representations is presented and very simple but
relatively precise expressions for the matrix elements are derived in both
representations. All analytical expressions for the quasiclassical radial
matrix elements in both representations contain only one special function --
the Anger function and its derivative. They generalize and increase the
accuracy of some known quasiclassical expressions. The small difference
between the two forms of the expressions for the dipole matrix elements
indicates to the applicability of the simple expressions given by the
consistent quasiclassical approach even for low atomic states.

It is important to note that the dipole matrix elements {\it as the
analytical functions} (even for the hydrogenic atom) {\it are expressed
through the Anger but not through the Bessel functions.} It is another thing
that the Anger functions ${\bf J}_\nu \left( z\right) $ of the integer order
$\nu =m$ coincide with the Bessel functions $J_m(z)$, i.e. ${\bf J}_m\left(
z\right) =(-1)^m{\bf J}_{-m}\left( z\right) =J_m(z).$ Expression of the
dipole matrix elements through the Bessel function $J_s(es)$ or through the
Anger function of the positive order and positive argument ${\bf J}_s\left(
es\right) $ (see, e.g. Delone e{\it t al} 1994) results to the erroneous
limit when $s\rightarrow 0$ and to another inaccuracies. \vspace{1cm} \\
{\bf Acknowledgments }\vspace{0.5cm} \\ The research described in this
publication was made possible in part by Grant No. LHV100 from the Joint
Fund Program of Lithuanian Government and International Science Foundation.
The author is also indebted to the referee for the useful comments and
suggestions for the improvement of this work.

\newpage {\bf References }\vspace{0.2cm}

\begin{description}
\item  Bethe H A and Salpeter E E 1957 {\it Quantum Mechanics of One- and
Two- Electron Atoms} (Berlin: Springer)

\item  Crossley R J S 1969 {\it Adv. At. Mol. Phys}. {\bf 5} 237--296

\item  Davydkin V A and Zon B A 1981 {\it Sov. Phys.-Opt. Spectrosc.} {\bf 51
} 13-5

\item  Delone N B, Goreslavsky S P and Krainov V P 1989 {\it J. Phys. B: At.
Mol. Opt. Phys}. {\bf 22} 2941--5

\item  --- 1994 {\it J. Phys. B: At. Mol. Opt. Phys}. {\bf 27} 4403--19

\item  D'yachkov L G and Pankratov P M 1991 {\it J. Phys. B: At. Mol. Opt.
Phys}. {\bf 24} 2267--75

\item  --- 1994 {\it J. Phys. B: At. Mol. Opt. Phys}. {\bf 27} 461--72

\item  Goreslavsky S P, Delone N K and Krainov V P 1982 {\it Sov. Phys.-JETP}
{\bf 55} 1032

\item  Gontis V and Kaulakys B 1987 {\it J. Phys. B: At. Mol. Phys}. {\bf 20}
5051--64

\item  Heim T M, Trautmann D and Baur G 1989 {\it J. Phys. B: At. Mol. Opt.
Phys}. {\bf 22} 727--40

\item  Kaulakys B 1991 {\it J. Phys. B: At. Mol. Opt. Phys}. {\bf 24} 571--85

\item  Kaulakys B and Vilutis G 1995 in {\it Chaos - The Interplay between
Stochastic and Deterministic Behaviour, }eds. P Garbaczewski{\it , }M Wolf
and A Weron, Karpacz'95 Proc, {\it Lecture Notes in Physics} Vol. 457
(Springer-Verlag,) p. 445-50

\item  Landau L D and Lifshitz E M 1985 {\it Quantum Mechanics} (New York:
Pergamon)

\item  Nana E S G, Owono O L C, Dada J P, Waha N L, Kwato N M G, Oumarou B
and Motapon O 1995 {\it J. Phys. B: At. Mol. Opt. Phys}. {\bf 28 }2333-53

\item  Pankratov P and Meyer-ter Vehn J 1992 {\it Phys. Rev }A {\bf 46}
5500--5
\end{description}

\end{document}